\documentclass[nolinenumbers]{aastex631}

\usepackage{amsmath}
\usepackage{graphicx}	

\newcommand{\tianshu}[1]{{#1}}

\begin{document}

\title[Lowest]{Supernova Explosions of the Lowest-Mass Massive Star Progenitors}

\author[0000-0002-0042-9873]{Tianshu Wang}
\affiliation{Department of Astrophysical Sciences, Princeton University, Princeton, NJ 08544}

\author[0000-0002-3099-5024]{Adam Burrows}
\affiliation{Department of Astrophysical Sciences, Princeton University, Princeton NJ 08544 and the Institute for Advanced Study, 1 Einstein Drive, Princeton NJ  08540}

\correspondingauthor{Tianshu Wang}
\email{tianshuw@princeton.edu}




\date{Accepted XXX. Received YYY}

\begin{abstract}
We here focus on the behavior of supernovae that technically explode in 1D (spherical symmetry). When simulated in 3D, however, the outcomes of representative progenitors of this class are quite different in almost all relevant quantities. In 3D, the explosion energies can be two to ten times higher, and there are correspondingly large differences in the $^{56}$Ni yields. These differences between the 3D and 1D simulations reflect in part the relative delay to explosion of the latter and in the former the presence of proto-neutron star convection that boosts the driving neutrino luminosities by as much as $\sim$50\% at later times. In addition, we find that the ejecta in 3D models are more neutron-rich, resulting in significant weak r-process and $^{48}$Ca yields. Furthermore, we find that in 3D the core is an interesting, though subdominant, source of acoustic power. In summary, we find that though a model might be found theoretically to explode in 1D, one must perform supernova simulations in 3D to capture most of the associated observables. The differences between 1D and 3D models are just too large to ignore.
\end{abstract}


\keywords{(stars:) supernovae: general -- (stars:) neutron -- (stars:)  -- hydrodynamics}



\section{Introduction}
\label{intro}


The emerging theory of core-collapse supernova explosions (CCSNe) suggests that there is most often a delay in explosion after the bounce of the Chandrasekhar-like core of most massive-star CCSN progenitors \citep{Janka2012,Burrows2021}. During this delay, turbulence driven by neutrino heating behind the shock in the so-called gain region \citep{Bethe1985} builds and becomes more vigorous over time, while the mass accretion rate through the stalled shock slowly ebbs.  At some critical point \citep{goshy}, either when the silicon-oxygen shell and its associated density jump is accreted and/or when the mantle between the accretion shock and the inner core otherwise becomes unstable and can't maintain quasi-hydrostatic equilibrium \citep{wang2022}, the shock is launched. The basic agencies of explosion are continuing neutrino heating and the turbulent hydrodynamic stress behind the shock and for most progenitors the turbulence is crucial to the explosion \citep{Burrows1995,janka1996,wang2022,burrows2024}. The accumulated energy in the post-shock ejecta, when corrected for the binding energy of the outer stellar mantle, starts negative and builds over time, reaching its asymptotic positive value sometime between $\sim$0.5 and $\sim$10 seconds after bounce. The timing of explosion and its subsequent development depend centrally on the mass density profile of the Chandrasekhar-like core at the time of the collapse instability, frequently associated with a ``compactness" parameter \citep{oconnor2011} \footnote{Defined as $\frac{M/M_\odot}{R(M)/1000\text{km}}$, we have generally set $M$ equal to 1.75 $M_{\odot}$. However, for the compact structures encountered at the lowest supernova progenitor masses, calculating the compactness at such large interior masses makes little sense. Hence, on Table 1 we also provide the initial central density and entropy, quantities better able to distinguish such models.} which increases roughly, though not monotonically \citep{Sukhbold2016,Sukhbold2018}, with progenitor mass. The lower compactness progenitors generally seem to explode with lower energies, lower $^{56}$Ni yields, lower recoil kicks, and more spherically.  When higher compactness models explode, they do so with greater explosion energies, greater $^{56}$Ni yields, greater asphericities, and larger kicks (though the latter with significant scatter; \citet{burrows_kick_2023}). While there is not yet quantitative agreement among supernova theorists performing sophisticated 3D simulations \citep{Lentz2015,roberts2016,Takiwaki2016,Muller2017,ott2018_rel,Glas2019,muller_low_kick2019,Burrows2019,Burrows2020,Muller2020,Powell2020,Stockinger2020,Kuroda2020,Bollig2021,Vartanyan2021,Vartanyan2023,burrows2024}, there is overall qualitative agreement on the specifics articulated here.  \citet{burrows2024} summarize the systematic correlations between observables, compactness, and progenitor mass seen in their comprehensive suite of state-of-the-art 3D simulations. 

However, the 8.8 $M_{\odot}$ (n8.8) solar-metallicity progenitor of \citet{Nomoto1984} and \citet{nomoto1988} and both the 8.1 $M_{\odot}$ (u8.1; $10^{-4}$ solar metallicity) and 9.6 $M_{\odot}$ (z9.6; zero-metallicity) models (A. Heger, private communication), with the lowest compactnesses published, do explode in 1D \citep{Kitaura2006,burrows2007_after,janka2008,fischer2010,hudepohl2010,melson2015b}. Though this is the realm of the so-called ``ONe," or ``electron-capture" progenitors \citep{Nomoto1984,nomoto1988,leung2020,limongi2024}, the extremely-low metallicity models u8.1 and z9.6 in fact have iron cores at the time of collapse; however, all such models (including n8.8) have NSE (nuclear statistical equilibrium) core compositions at bounce. For such models that explode in 1D the turbulent stress and 3D effects are not crucial factors in the inauguration of explosion. This is because their rarified envelopes lead to a post-bounce mass accretion rate that plummets immediately after bounce, and with it the accretion ram pressure, while the early post-bounce neutrino luminosity (most of it sourced by the diffusive component from the shocked proto-neutron star (PNS) inner mantle) is adequate to power a prompt, neutrino-driven explosion\footnote{For more massive progenitors with shallower initial density profiles, post-bounce accretion contributes significantly to the driving neutrino luminosities themselves.}, though with low asymptotic supernova energy, and such explosions transition into a neutrino-driven wind very early on \citep{wang2023}.


\begin{figure}
    \centering
    \includegraphics[width=0.45\textwidth]{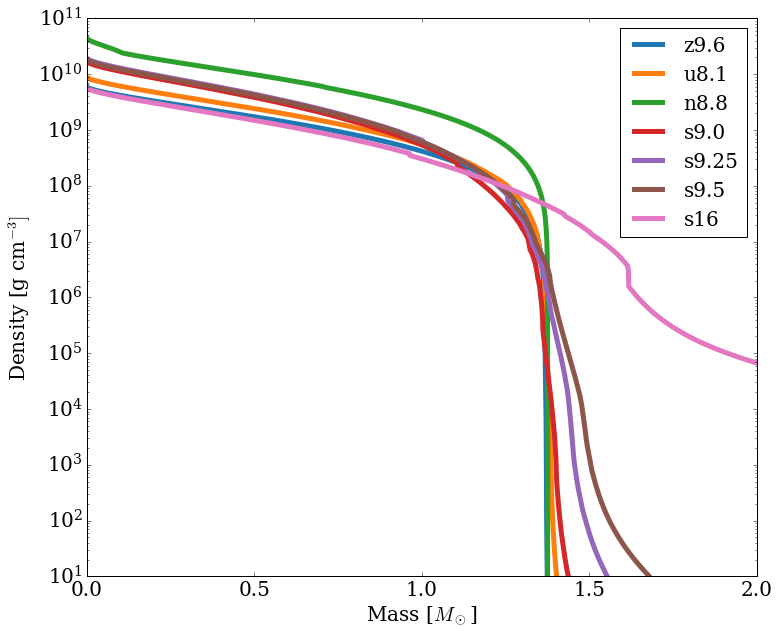}
    \includegraphics[width=0.45\textwidth]{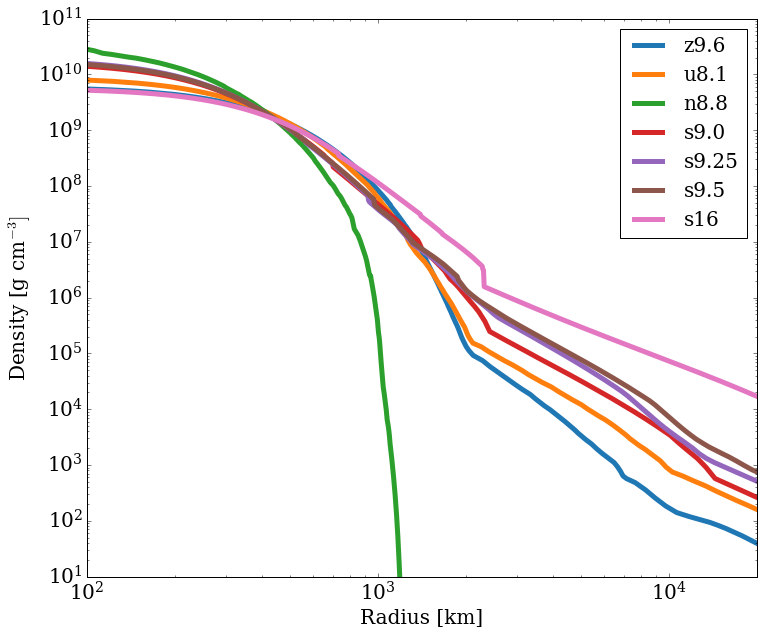}    
    \caption{Progenitor mass density profiles as a function of interior mass coordinate (left) and radius (right) for the extremely low compactness, low or zero metallicity, models of this study (u8.1, z9.6), compared with the corresponding profiles of low-mass solar-metallicity progenitors (\tianshu{s9.0a, s9.0b}, s9.25, s9.5) \tianshu{and an example of a higher-mass progenitor (s16)}. The 8.8 solar-metallicity progenitor \citep{Nomoto1984,nomoto1988} is also plotted. The \tianshu{s9.0a, s9.0b}, s9.25, s9.5, and s16 comparison models are from \citet{Sukhbold2016} and \citet{Sukhbold2018} and the extreme compactness models (u8.1 and z9.6) that explode in 1D are from A. Heger (private communication). The s9.0 model does not seem to explode in 1D. See text for a discussion.} 
    \label{fig:M-rho}
\end{figure}

\citet{Nomoto1984} introduced the idea that ONe cores (such as his 8.8 $M_{\odot}$ model) achieve the effective Chandrasekhar mass and start to collapse not by the photodissociation of iron-peak nuclei, but by electron capture, and that there is burning on infall.  However, what type of inner structures the stars in the $\sim$8$-$9 $M_{\odot}$ progenitor mass range (and for a spread of metallicities) actually possess at the end of their lives has not been definitively determined.  This is in part due to the many thermal pulsations such cores experience and their electron degeneracy.  The latter can lead to many flashes that are difficult to simulate, and the former significantly retard progress to collapse in extant stellar evolution codes. The result is not only ambiguity, but that there are very few published models in this mass range available to supernova modelers. \citet{limongi2024} and \citet{leung2020} discuss these issues (see also \citet{jones2019}) and \citet{limongi2024} suggest that solar-metallicity stars in the ZAMS mass range $\sim$8.2$-$9.2 $M_{\odot}$ will be electron-capture supernovae. However, due to myriads of intervening thermal pulses at the end of their calculations they have yet actually to reach the mass density threshold for electron capture on $^{20}$Ne. 

Nevertheless, despite the remaining uncertainties concerning the core structures of massive stars in the 8$-$9 $M_{\odot}$ ZAMS mass range, the existence of theoretical evolutionary models that lead to explosions in 1D (spherical) enables a study of the effects of 3D through a comparison of 1D and 3D simulations. This is the subject of this paper. In addition, the Salpeter mass function indicates that this range of ZAMS mass space occupies a disproportionate fraction of progenitor space, though remaining uncertainties in stellar evolution leave the actual fraction a subject of conjecture. Should all stars between 8 and 9 $M_{\odot}$ technically explode in 1D, then this would constitute $\sim$15\% of massive star deaths and nearly 15\% of core-collapse supernovae.

\begin{figure} 
    \centering
    \includegraphics[width=0.48\textwidth]{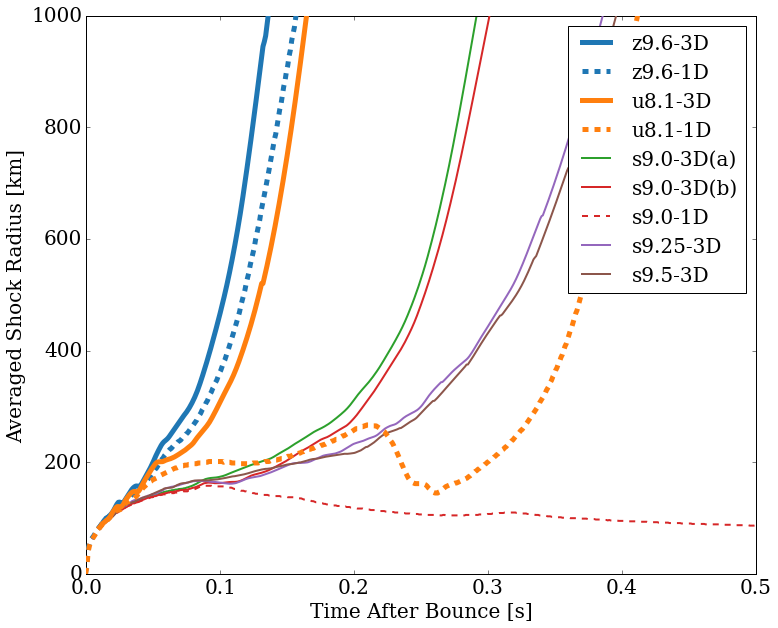}
    \includegraphics[width=0.48\textwidth]{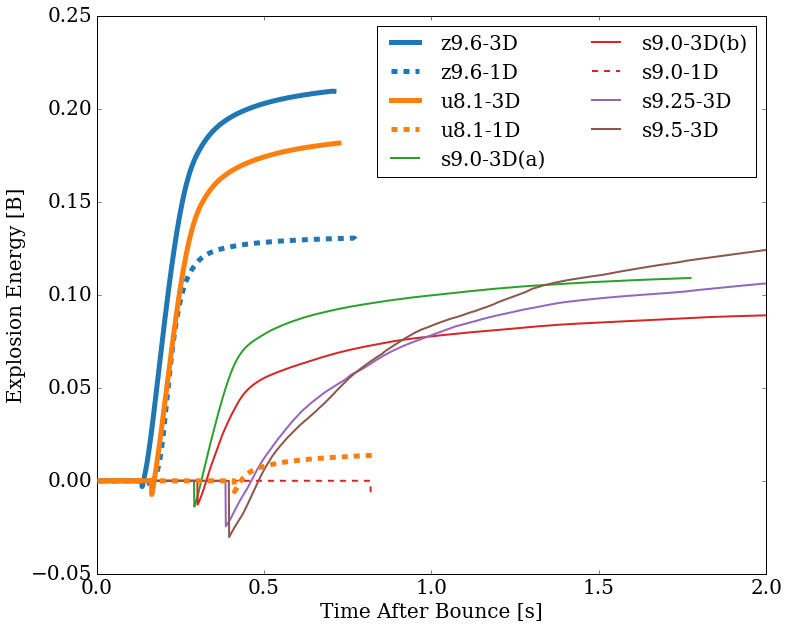}    
    \caption{The temporal evolution of the average shock radius (left) and the explosion energy (right, in Bethes $\equiv$10$^{51}$ ergs) for 3D models (solid) and 1D models (dashed, for only the u8.1, z9.6, and s9.0 progenitors). Note that the 1D s9.0 model does not explode, while the corresponding 1D models for the u8.1 and z9.6 models do. Note also that the 3D u8.1 ($10^{-4}$ solar) and z9.6 (zero metallicity) models explode with a greater asymptotic energy than the 3D solar-metallicity models s9.0, s9.25, and s9.5. Finally, we call the reader's attention to the significant relative delay in the explosion of the 1D u8.1 model and its very low asymptotic explosion energy. See text in \S\ref{hydro} for a discussion.}
    \label{fig:t-rs-Mpns}
\end{figure}

\begin{figure} 
    \centering
    \includegraphics[width=0.96\textwidth]{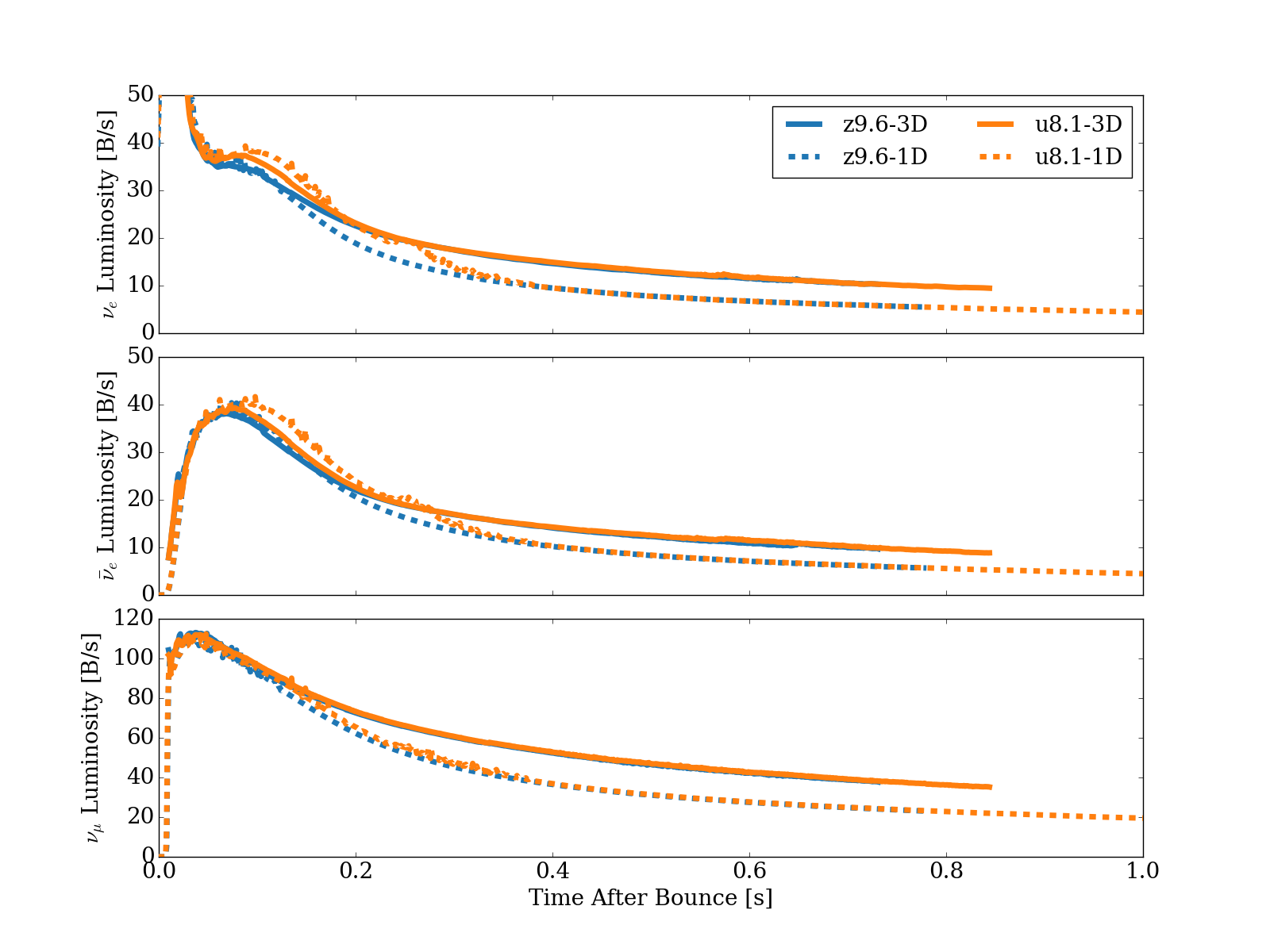}
    \includegraphics[width=0.48\textwidth]{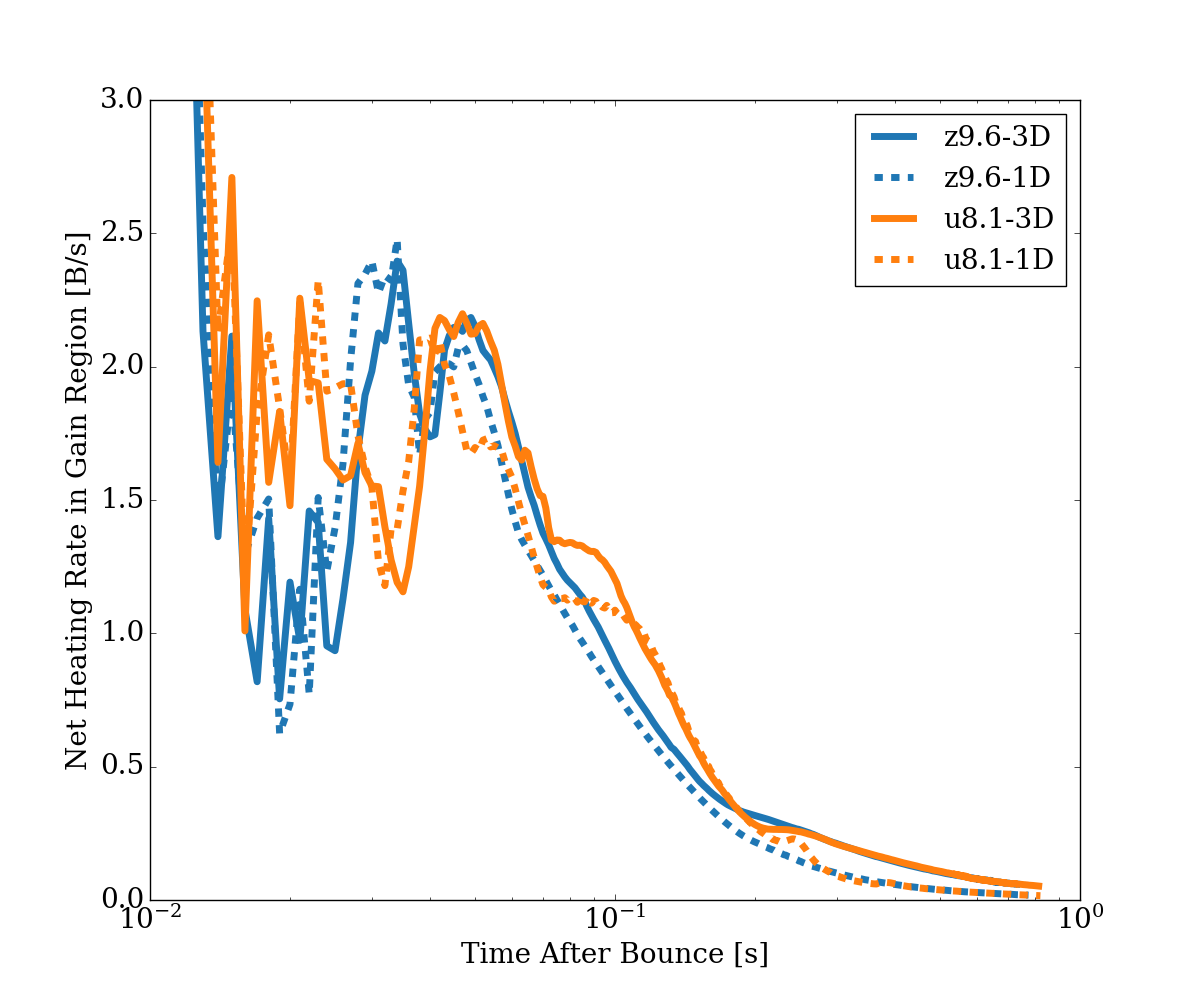}
    \includegraphics[width=0.48\textwidth]{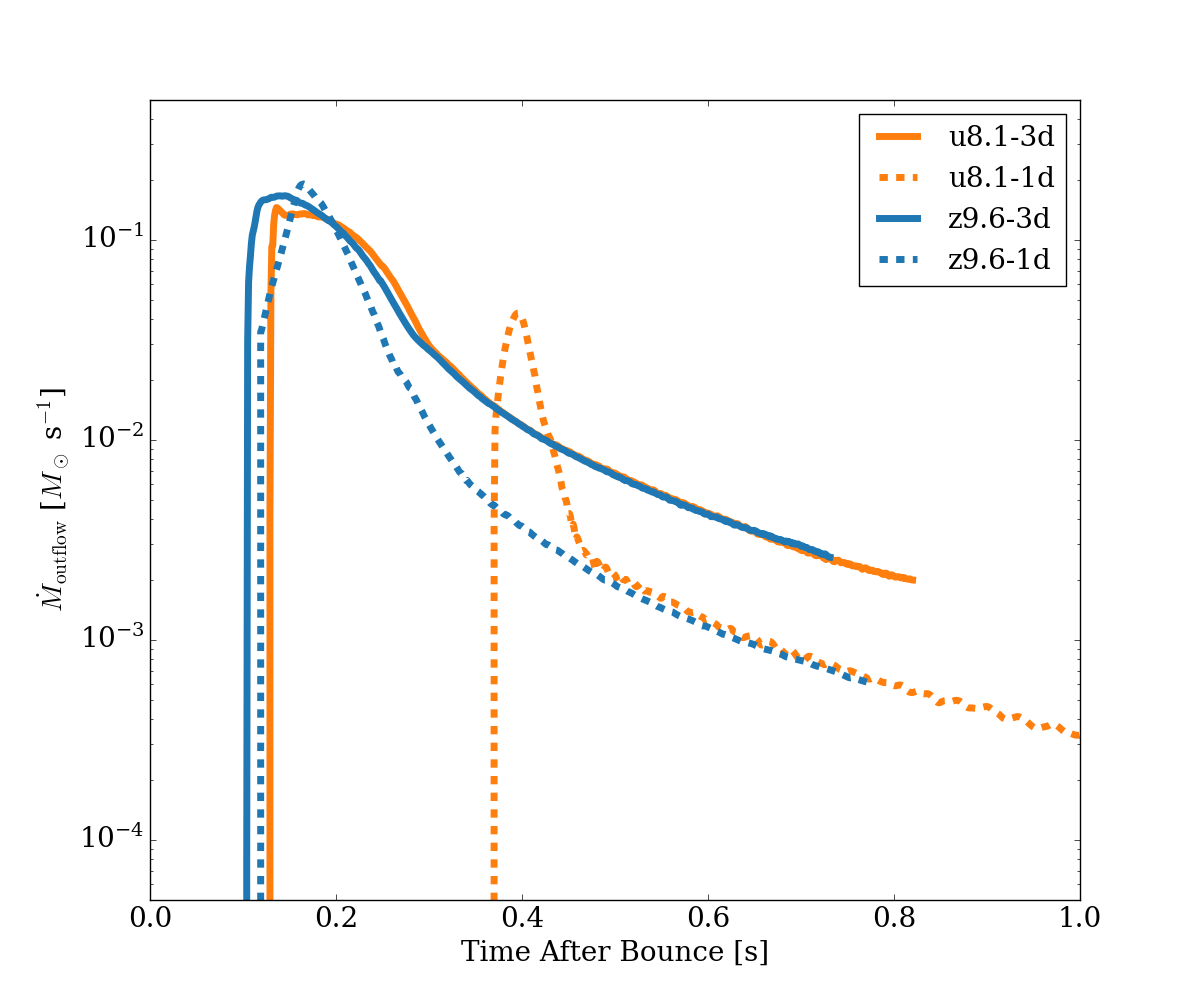}
    \caption{The temporal evolution of the neutrino luminosities (Top) for the u8.1 (orange) and z9.6 (blue) models in both 1D and 3D.  The net neutrino heating rate evolution (note the log$_{10}$ time scale) in the gain region is depicted in the bottom left. PNS convection enhances the neutrino luminosities of all species at later times and leads to higher net heating rates (bottom left). This in turn elevates the neutrino-driven wind mass loss rate (bottom right) in 3D vis \`a vis 1D. The higher neutrino luminosities in 1D at early times are due to the later explosion times of such models and the larger consequent accretion rates during this more extended pre-explosion phase.  This effect is much larger for the u8.1 model than for the z9.6 model.}  
    \label{fig:neutrino}
\end{figure}

\begin{figure} 
    \centering
    \includegraphics[width=0.48\textwidth]{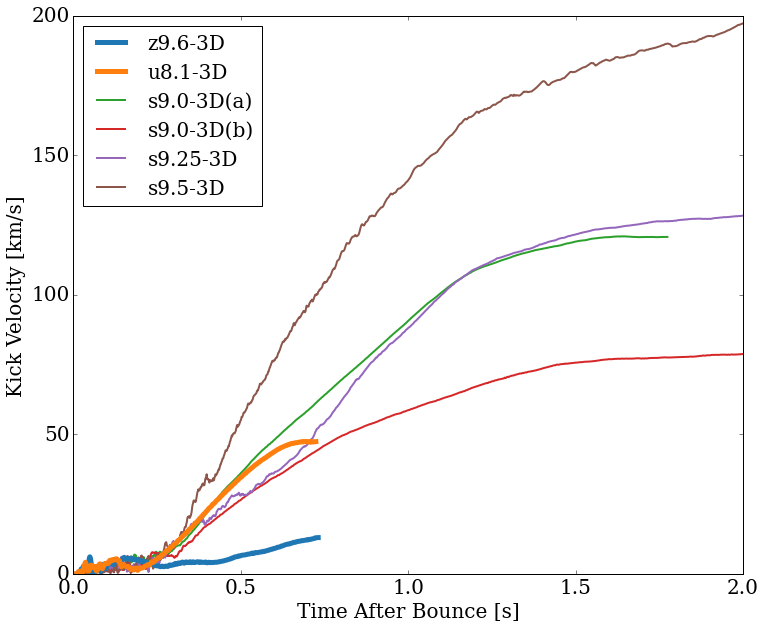}
    \caption{The temporal evolution of the kick speeds for the 3D models compared in this study. The recoil kicks for the extreme models u8.1 and z9.6 are quite low and are due predominantly to the neutrino contribution. See text in \S\ref{hydro} and \citet{burrows_kick_2023} for a discussion of the broader context.}
    \label{fig:t-rs-kick}
\end{figure}

We begin in \S\ref{hydro} by describing the distinctive hydrodynamics of 1D, prompt, neutrino-driven explosions and compare with the same models in 3D. We also summarize their explosion energies, the recoil kick speeds, and the gravitational mass range of their residues. In \S\ref{sound}, we discuss sound wave generation and the characteristics of the acoustic component and in \S\ref{nucleo} we provide the distinctive nucleosynthesis. The latter includes the $^{56}$Ni, $^{57}$Ni, and $^{44}$Ti, and $^{48}$Ca yields, as well as their possible weak r-process production. We conclude with a general discussion of the results and thoughts concerning future theoretical directions.

\section{General Hydrodynamics and Results}
\label{hydro}




The radiation/hydrodynamic code we use for this study is F{\sc{ornax}} \citep{Skinner2019,vartanyan2019a,Burrows2020,burrows2024}. For the 1D simulations, we use a radial grid with 1024 zones and for the 3D simulations we use a 1024($r$)$\times$128($\theta$)$\times$256($\phi$) grid. The number of neutrino energy groups is twelve (for $\nu_e$, $\bar{\nu}_e$, and ``$\nu_{\mu}$" neutrinos, where the latter bundles $\nu_{\mu}$, $\bar{\nu}_{\mu}$, ${\nu}_{\tau}$, and $\bar{\nu}_{\tau}$ neutrinos).
We employed the SFHo equation of state (EOS) for all simulations \citep{Steiner2013}. 

\tianshu{We use the u8.1 ($10^{-4}$ solar) and z9.6 (zero metallicity) progenitors from A. Heger (private communication) as example models that explode in 1D. These progenitors, other than the Nomoto 8.8 $M_{\odot}$ model, have the lowest published compactness. We compare these low-mass models with slightly higher-compactness models that explode only in 3D (s9.0a, s9.0b, s9.25, and s9.5, all of which have solar metallicity \citep{Sukhbold2016}). Models s9.0a and s9.0b differ only in the imposition of perturbations on infall in the former, with consequences predominantly for the nucleosynthesis (see \citet{wang2023,wang2023b}). All these comparison models have already been published \citep{wang2023,wang2023b,burrows2024}. The outer boundary radius is 20,000 kilometers (km) for models u8.1 and z9.6, 30,000 km for models s9.0a, s9.0b and s9.5, and 45,000 km for model s9.25.}


\tianshu{Figure \ref{fig:M-rho} depicts the initial density profiles of all models mentioned above, together with the solar-metallicity 8.8 $M_\odot$ progenitor \citet{nomoto1988} that also explodes in 1D (labeled as n8.8) and a higher-mass comparison progenitor (16 $M_\odot$ with solar metallicity, labeled as s16). It is clear that models that explode in 1D (z9.6, u8.1, and n8.8) have lower densities in their mantles and have the steepest mantle density gradients. Since the density decreases rapidly at interior masses of 1.3$-$1.5 $M_\odot$,} calculating the compactness at 1.75 $M_{\odot}$, as we normally do, may not capture the extreme contrasts between core and mantle for those models that explode in 1D. Therefore, it may not be as useful a discriminant. Though for those models the compactness is extremely low, using the central density or entropy at collapse may be better \citep{temaj2023} and we provide these quantities in summary Table 1. However, for now we will not focus on specific discriminating indices. Rather, we seek to explore the differences between 1D and 3D developments when the 1D models explode and leave to future work the detailed \tianshu{classification of the core structures of the least massive CCSN progenitor stars.}

Figure \ref{fig:t-rs-Mpns} portrays the mean shock radius versus time after bounce for the 1D (dashed) and 3D (solid) models (left panel) and also provides for comparison the corresponding development of the 3D models s9.0a, \tianshu{s9.0b}, s9.25, and s9.5. 
We note that though the 1D u8.1 model explodes, that explosion is delayed significantly.  Its explosion is marginal, with an energy of only $\sim$10$^{49}$ ergs (see Table 1). This is a direct consequence of its slightly thicker envelope than found for the z9.6 model (Figure \ref{fig:M-rho}).  A slightly enhanced initial mantle density translates into a slightly greater mass accretion rate ($\dot{M}$) and accretion ram pressure after bounce.  These delay the explosion until the accretion rate subsides. \tianshu{Since they are not aided by important 3D effects, such as neutrino-driven turbulence, in overcoming the ram pressure, one-dimensional models suffer greater delays.} For many progenitor models, such an eventual decrease in accretion rate is accompanied by a corresponding decrease in driving luminosity, with the result that explosion is not assured \citep{wang2022}, \tianshu{and this is why the s9.0-1D model in Figure \ref{fig:t-rs-Mpns}, and most other 1D models, aren't seen to explode.} However, for the models that explode in 1D, the diffusive component of the luminosity provides a floor, so that the rapid decrease in $\dot{M}$ is not accompanied by a decrease in the driving neutrino luminosities.  The result is explosion, though the asymptotic explosion energy (Figure \ref{fig:t-rs-Mpns}, right panel), contingent as it is upon the integrated mantle neutrino heating rate, is generically low.

As the right-hand-side of Figure \ref{fig:t-rs-Mpns} demonstrates (and Table 1 summarizes), the final explosion energies for the z9.6 model are $\sim$0.125 Bethes (1D) and $\sim$0.22 Bethes (3D), while for the u8.1 model they are $\sim$0.017 Bethes (1D) and $\sim$0.17 Bethes (3D)\footnote{One Bethe (B) is 10$^{51}$ ergs.}. Interestingly, the explosion energies in 3D for the u8.1 and z9.6 models are slightly higher than the explosion energies for the s9.0, s9.25, and s9.5 progenitors, despite their larger compactnesses and neutrino luminosities.
It is the greater delay to explosion and the associated wasting of the heat deposited during a longer stall to explosion of the s9.0, s9.25, and s9.5 progenitors (despite their slightly higher luminosities at the same post-bounce times) that results in slightly lower asymptotic explosion energies for them. This provides an ``asterisk" to the conclusion that explosion energy tracks compactness \citep{burrows2024} and is likely a metallicity effect $-$ published lower metallicity progenitors have higher compactnesses upon collapse \citep{Sukhbold2016}.

Figure \ref{fig:neutrino} portrays the evolution of the neutrino luminosities
of species $\nu_e$, $\bar{\nu}_e$, and ``$\nu_{\mu}$" of models z9.6 and u8.1 in both 1D and 3D.  We note that for the u8.1 model the $\nu_e$ and $\bar{\nu}_e$ luminosities in 1D start larger than the corresponding luminosities for the 3D model. This is due to the significant delay to explosion of this 1D model and the resultant larger mass accretion rate (and corresponding accretion luminosity) at this earlier phase compared with that found in 3D, which explodes much earlier.  As Figure \ref{fig:M-rho} shows, the envelope of the u8.1 model is slightly thicker than that for the z9.6 model. As seen in Figure \ref{fig:t-rs-Mpns}, this results \tianshu{in} a quicker explosion in 1D for the latter than the former.  

However, at slightly later times, the luminosities of the 3D models overtake those for the 1D models for both the z9.6 and u8.1 progenitors and continue higher thereafter. This is a direct consequence of proto-neutron star (PNS) convection in the residual core and is a purely multi-D effect.  Driven by negative lepton gradients in the PNS \citep{Dessart2006,Nagakura2019}, such convection dredges heat from the core, is found during the first second after bounce in these models between about 15 and 30 km, and continues unabated for the duration of the PNS cooling phase.

The left lower panel of Figure \ref{fig:neutrino} shows the net neutrino heating rate in the gain regions for the 1D and 3D models of the u8.1 and z9.6 progenitors. Due to PNS convection and a slight physical expansion in the shock radius (and, hence, the gain volume) in 3D due to turbulence behind the initially decelerating shock, the neutrino heating rates in 3D are generally larger. As a result, the explosion energies in 3D are larger than in 1D.  This is particularly the case for the u8.1 model due to the significant delay in its 1D explosion. The difference in the wind mass loss rates between the 1D and 3D models (bottom right of Figure \ref{fig:neutrino}) at later times is also quite pronounced, amounting to approximately a factor of three in $\dot{M}$. This is a major difference in the behavior of 3D models and emphasizes that, though these models explode in 1D, one must perform 3D simulations to capture their true behavior. 

Figure \ref{fig:t-rs-kick} shows the evolutions of the \tianshu{PNS} recoil kick imparted to the 3D u8.1 and z9.6 models, and compares them to similar curves for the s9.0a,b, s9.25, and s9.5 models. A much more comprehensive discussion of this general topic can be found in \citet{burrows_kick_2023}. We see in Figure \ref{fig:t-rs-kick} that the kicks are low, near $\sim$40 km s$^{-1}$ for the u8.1 model and at $\sim$13 km s$^{-1}$ for the z9.6 model.  Not shown on this plot is that these modest kicks are due mostly to anisotropic neutrino emissions \tianshu{\citep{vartanyan2019}}. The ejecta are mostly launched spherically and do not contribute to the same degree. That this is likely the case for the lowest mass progenitors was discussed and motivated in \citet{burrows_kick_2023}.

We provide the final neutron star baryon and gravitational masses for the residual neutron stars in Table 1, along with a variety of other summary model characteristics and results. We are using the SFHo EOS and note that the gravitational mass in particular is contingent upon this.  Nevertheless, we find gravitational masses for the u8.1 and z9.6 neutron stars near a low value of $\sim$1.25 $M_{\odot}$. Note that the gravitational mass for the residue of the s9.0 model is expected to be slightly lower still, as is the neutron star mass for the residues of accretion-induced collapse \citep{dessart2007}. 

\begin{figure} 
    \centering
    \includegraphics[width=0.96\textwidth]{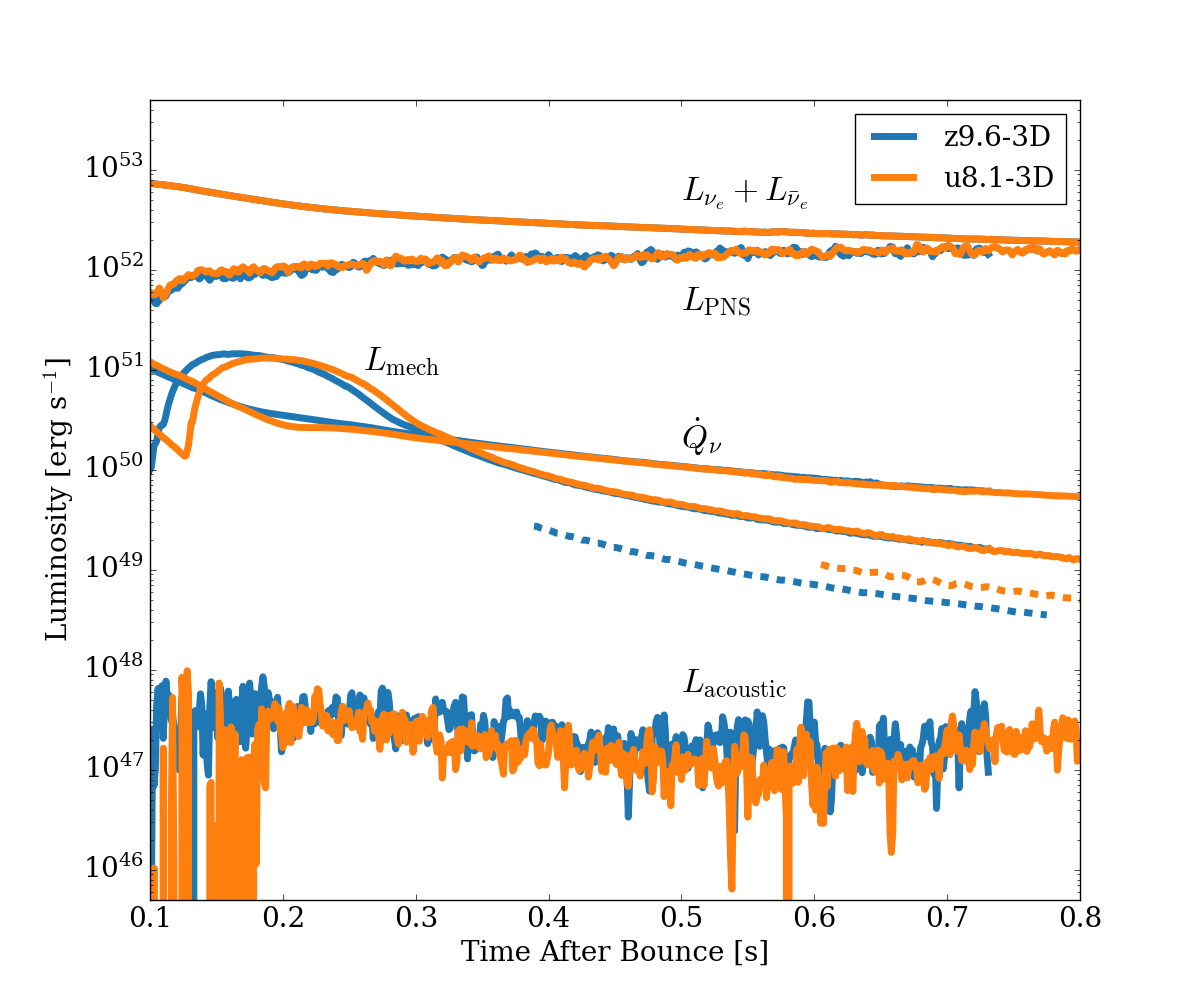}
    \caption{The neutrino ($L_{\nu_e} + L_{\bar{\nu}_e}$), mechanical, PNS convective, and acoustic luminosities of the z9.6 and u8.1 models in the neutrino-driven wind phase. The net neutrino heating rate ($\dot{Q}$) for is also shown. The mechanical luminosity is measured at 500 km, the convective luminosity is the maximum value in the PNS convective region, and the acoustic luminosity is measured just above the PNS radius defined at 10$^{11}$ g cm$^{-3}$. Note that the plots start at 0.1 seconds after bounce. Also shown for comparison is \tianshu{$L_{\rm mech}$} for the 1D models (dashed) at later times, well into the wind phase.}  
    \label{fig:acoustic-power}
\end{figure}

For models that explode in 1D (and their 3D counterparts), the hydrodynamics transitions quickly into a wind phase \citep{duncan1986,burrows1987,Burrows1995,qian1996,wang2023}. In fact, for them the power in the wind contributes substantially to the total energy of the blast. Figure \ref{fig:acoustic-power} shows on the same plot the evolution of the various powers and luminosities associated with the 3D u8.1 and z9.6 models. Included are the sum of the $\nu_e$ and $\bar{\nu}_e$ luminosities, the peak of the convective luminosity in the PNS (\tianshu{$L_{\rm PNS}$}), the heating rate in the gain region ($\dot{Q}$) (repeated from Figure \ref{fig:neutrino}), the mechanical luminosity of the ejecta (\tianshu{$L_{\rm mech}$}, which clearly emerges as a wind after $\sim$0.3 seconds), and the acoustic power that emerges from the core into the exploding mantle (see \S\ref{sound}). 
Note that the plot starts at 0.1 seconds after bounce. 

To calculate these quantities we have used the formulae:
\begin{equation}
\begin{split}
        L_{\rm mech} &= r^2\int d\Omega \rho v_r (\frac{1}{2}(v_r^2+v_\theta^2+v_\phi^2)+e+\Phi+P/\rho)\\
        L_{\rm PNS} &= r^2\int d\Omega \rho v_{\rm turb} (\frac{1}{2}(v_{\rm turb}^2+v_\theta^2+v_\phi^2)+e+\Phi+P/\rho)\\
        L_{\rm acoustic} &= r^2\int d\Omega (P-\langle P\rangle_t)(v_r-\langle v_r\rangle_t)\, ,
\end{split}
\label{acoustic_power}
\end{equation}
where $\Phi$ is the gravitational potential, $v_{\rm turb}=v_r-\frac{1}{4\pi}\int d\Omega v_r$, and $\langle A\rangle_t$ is the time-averaged value of $A$. \tianshu{Intuitively, $L_{\rm mech}$ is the flux of total energy, $L_{\rm PNS}$ is the energy flux carried by convection in the PNS (where the radial speed, $v_{\rm turb}$ is the speed relative to the mean radial motion), and $L_{\rm acoustic}$ is the energy carried by acoustic waves in the frame of the radial motion.} In our calculation of the acoustic power, we calculate the time average using a 20-ms sliding window (50 Hz). This filters out the influence of the expanding shock. Changing this frequency cut to any value between 25 and 80 Hz doesn't influence the measured acoustic power (especially at late times), since the acoustic component generally has frequencies above 80 Hz (\S\ref{sound}).

Figure \ref{fig:acoustic-power} is quite illuminating. First (as suggested in the bottom right panel of Figure \ref{fig:neutrino}), the mechanical power of the late-time wind is approximately seven and four times higher in 3D than 1D for the z9.6 and u8.1 progenitors, respectively. This is a direct consequence of the PNS-convection-boosted neutrino luminosities and heating rates. Second, $\dot{Q}$ and \tianshu{$L_{\rm mech}$} in both 1D and 3D soon follow in lock step in an approximate power-law relation, with power 1.8 to 1.6 ($L_{mech} \propto \dot{Q}^{1.8-1.6}$), only gradually/secularly decreasing with time. Third, while during the first second after bounce the neutrino luminosity clearly exceeds $L_{\rm PNS}$ (early on by an order of magnitude), the two approach one another at later times as the accretion phase abates and the PNS settles into its quasi-static Kelvin-Helmholtz phase. The fluctuations in $L_{\rm PNS}$ are clearly seen in Figure \ref{fig:acoustic-power}\footnote{We note that the 3D calculations we have performed over the last $\sim$5 years using F{\sc{ornax}} follow 1) to the center, 2) to late times, and 3) in 3D the full Kelvin-Helmholtz contraction phase of neutron star birth.  This is the only context in stellar evolution theory for which almost an entire stellar evolutionary phase, with nested convective and radiative zones, has been followed in 3D.}.

\begin{table}[]
\centering
\begin{tabular}{c|cccccccccc}
\hline
$M_\text{ZAMS}$ &Dimension  &Metallicity        & $\rho_{\rm c, ini}$  &$S_{\rm c, ini}$ & $t_{\rm pb, max}$    & $M_{\rm b, final}$ & $M_{\rm g, final}$  & $E_{\rm exp}$ & $^{56}$Ni              & $\vert v_{\rm kick}^{\rm total}\vert$  \\
{[}$M_\odot${]} &           &Solar Units         & [g cm$^{-3}$]         & [$k_b$/baryon] & {[}s{]}           & {[}$M_\odot${]}     & {[}$M_\odot${]}  & {[}B{]}          & {[}$10^{-2}M_\odot${]} & {[}km s$^{-1}${]}   \\ \hline
u8.1            &3D         & $10^{-4}$         & $8.61\times10^{9}$    & 0.842 & 0.844             & 1.363               & 1.250            & 0.171            & 0.572                       & 37.8            \\
u8.1            &1D         & $10^{-4}$         & $8.61\times10^{9}$    & 0.842 & 2.715             & 1.386               & 1.269            & 0.017            & 0.164                       &     N/A         \\ 
z9.6            &3D         & $0$               & $5.80\times10^{9}$    & 0.888 & 0.721             & 1.356               & 1.245            & 0.207            & 0.566                       & 13.0                \\
z9.6            &1D         & $0$               & $5.80\times10^{9}$    & 0.888 & 0.775             & 1.365               & 1.252            & 0.129            & 0.922                       &     N/A            \\    
s9.0(a)         &3D         & 1                 & $1.63\times10^{10}$   & 0.532 & 1.775             & 1.347               & 1.237            & 0.111            & 0.168                  & 120.7        \\
s9.0(b)         &3D         & 1                 & $1.63\times10^{10}$   & 0.532 & 1.950             & 1.348               & 1.238            & 0.0945           & 0.612                  & 78.6          \\
s9.25           &3D         & 1                 & $1.87\times10^{10}$   & 0.540 & 3.532             & 1.378               & 1.263            & 0.124            & 1.04                   & 140.1          \\ 
s9.5            &3D         & 1                 & $1.78\times10^{10}$   & 0.540 & 2.375             & 1.397               & 1.278            & 0.142            & 1.47                   & 208.6          \\
\end{tabular}
\label{tab:table1}
\caption{Summary of simulation outcome data and initial model characteristics. $\rho_{\rm c, ini}$ and $S_{\rm c, ini}$ are the central density and entropy of the progenitor model (when the maximum infall speed reaches 950 km $s^{-1}$). $t_{\rm pb, max}$ is the post-bounce duration of the simulation. {We stop the simulations when the shock radii reach about 20,000 km. The u8.1-1D model explodes later than the other models, and with lower energy, with the result that it has a significantly lower shock velocity. Thus, this simulation was run much longer than the others.} $M_{\rm b, final}$ is the PNS baryon mass at the end of each simulation, while $M_{\rm g, final}$ is the derived gravitational mass of the cold neutron star. $E_{\rm exp}$ is the explosion energy and $\vert v_{\rm kick}^{\rm total}\vert$ is the PNS recoil kick speed at the end of each simulation. For the lowest mass CCSNe, $^{56}$Ni production (here in units of 10$^{-2}$ $M_{\odot}$) is lower in the 3D simulations due to the larger amount of neutron-rich ejecta. The u8.1 model is a special case, since its 1D counterpart explodes so weakly that the total ejecta mass is much lower than in 3D. Therefore, u8.1-1D $^{56}$Ni production is lower even if the $^{56}$Ni mass fraction is higher than in 3D.  }
\end{table}

\section{Sound Generation and Acoustic Power}
\label{sound}

\begin{figure} 
    \centering
    \includegraphics[width=0.48\textwidth]{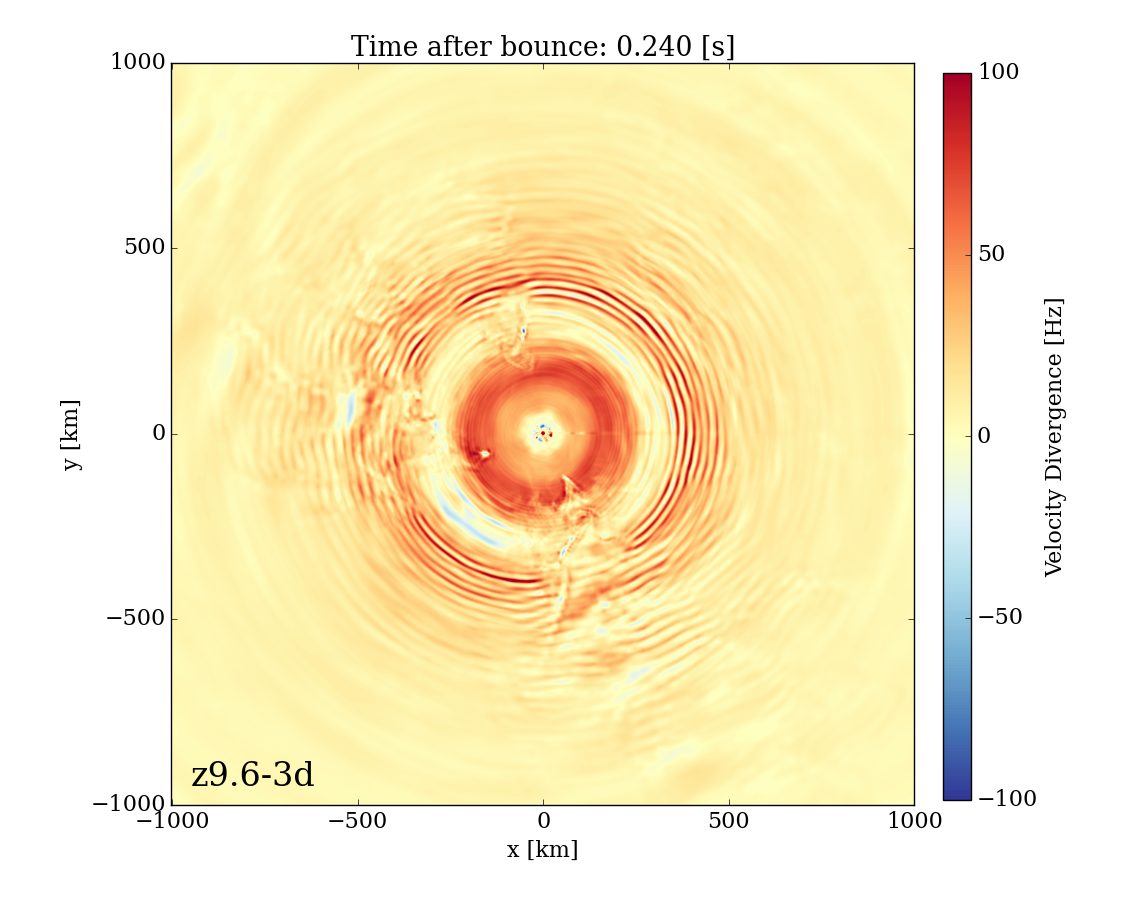}
    \includegraphics[width=0.48\textwidth]{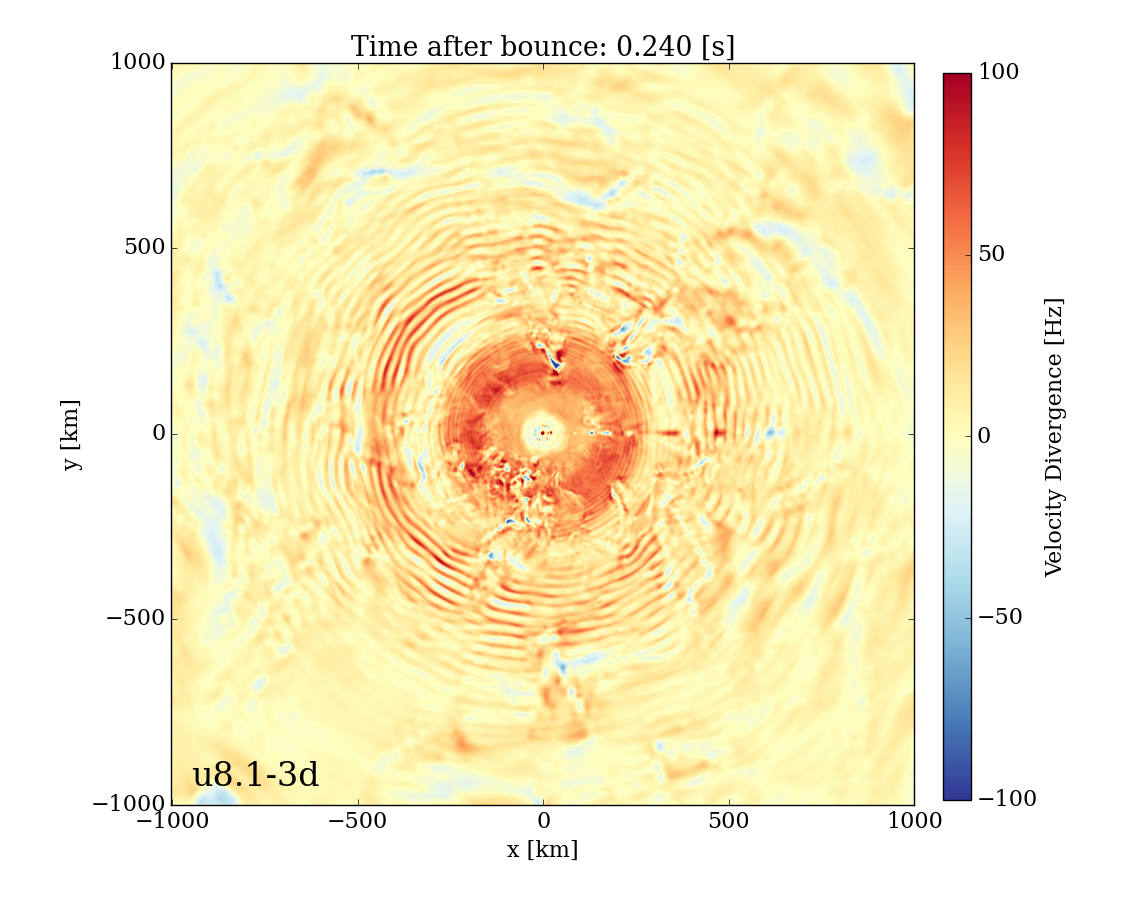}
    \includegraphics[width=0.48\textwidth]{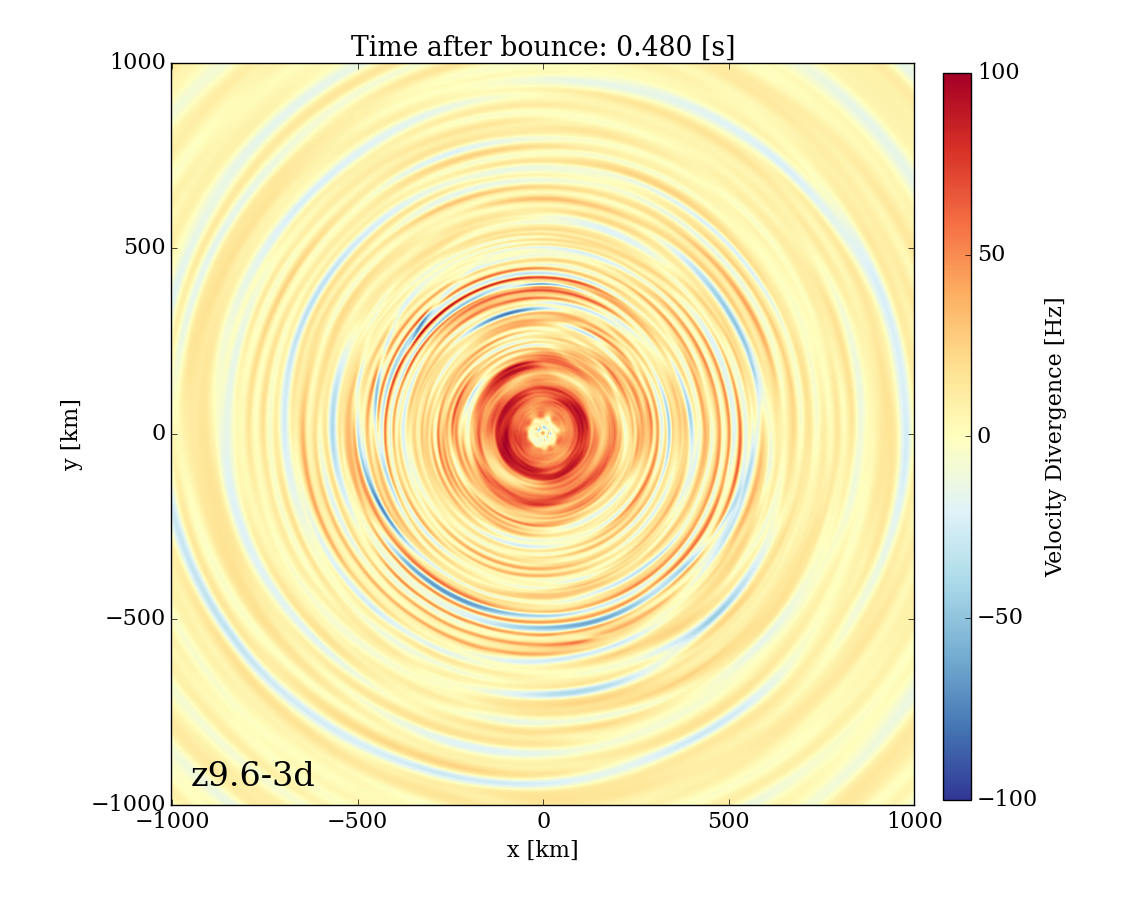}
    \includegraphics[width=0.48\textwidth]{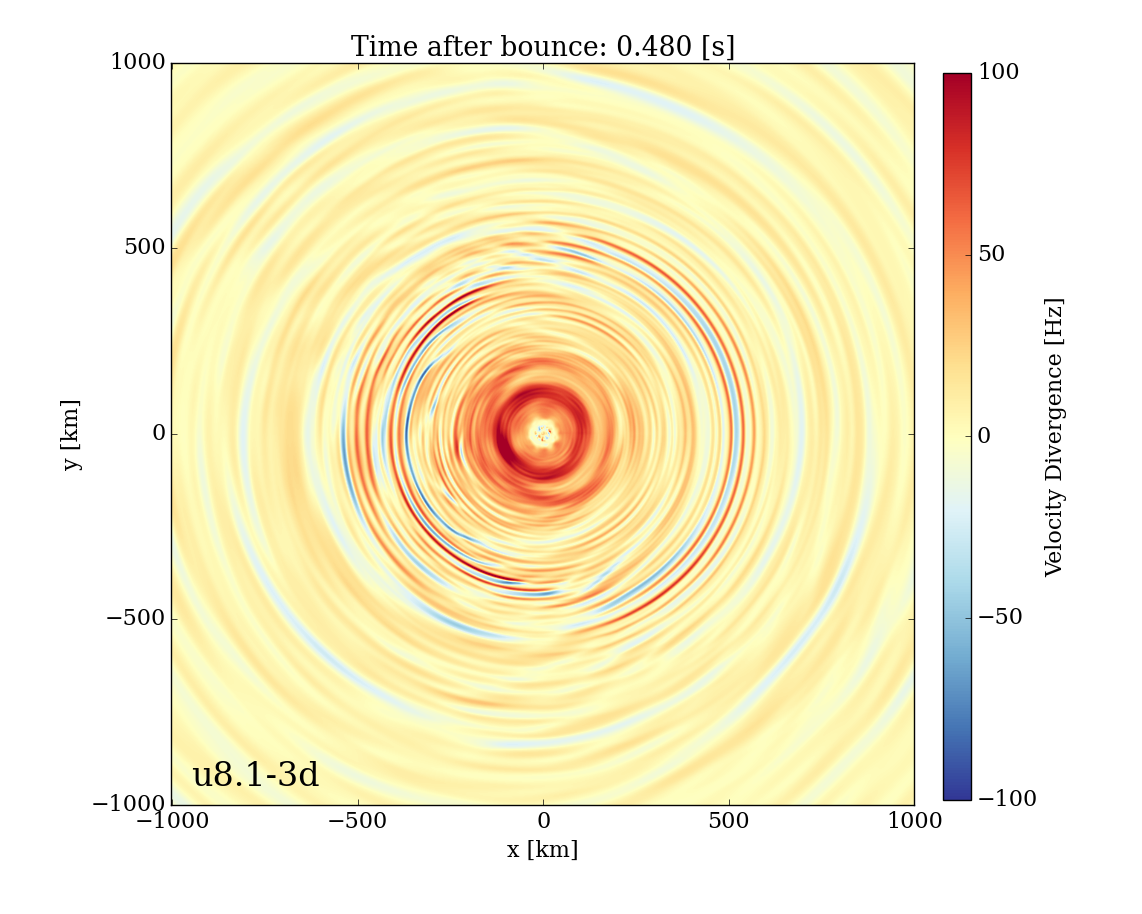}
    \caption{Velocity divergence at 240 ms and 480 ms post-bounce on the x-y plane of the 3D z9.6 and u8.1 models. The presence of sound waves generated in the interior is clear. See text in \S\ref{sound} for a discussion.}
    \label{fig:snapshot}
\end{figure}

\citet{nevins2023} have speculated that the nucleosynthetic yields of such lowest-compactness progenitors could be affected by this acoustic component. \citet{gossan} have suggested that acoustic power could help drive the supernova explosion. Figure \ref{fig:snapshot} depicts snapshots at two different times of the divergence of the velocity in the inner 1000 km of the explosions of the u8.1 and z9.6 3D models. Roughly concentric waves of velocity perturbations are clearly seen.  Movies of these waves show that these perturbations propagate at speeds of $c_s + v$, where $c_s$ is the local speed of sound
and $v$ is the wind speed. After inner turbulence settles (within $\sim$0.2 seconds of bounce), a Fourier analysis reveals that these sound waves have frequencies between roughly 100 and 250 Hertz (Hz). We comment that no such waves are seen in the 1D simulations.

\begin{figure} 
    \centering
    \includegraphics[width=0.96\textwidth]{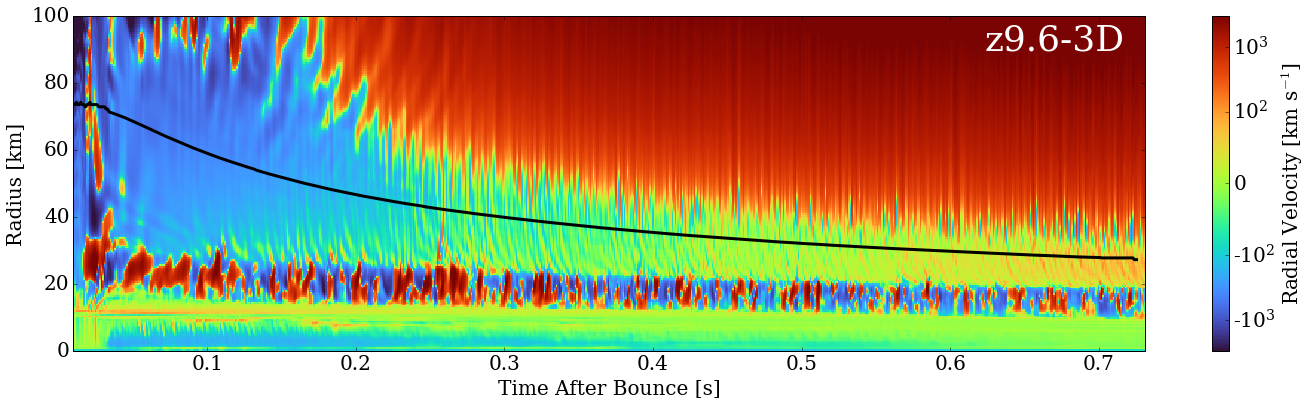}
    \includegraphics[width=0.96\textwidth]{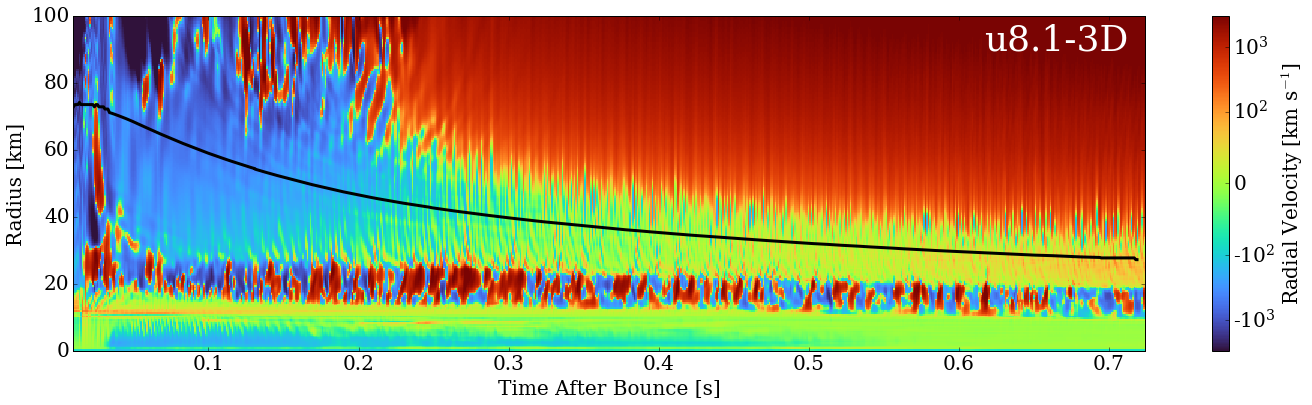}
    \caption{Space-time diagrams of radial velocity along $(\theta, \phi)=(\pi/2, 0)$ (along the x-direction of Figure \ref{fig:snapshot}). 
    Top: Radial velocity of the z9.6 model. Bottom: Same as the top panel, but for the u8.1 model. The solid black line shows the PNS radius (defined as the $\rho=10^{11}$g cm$^{-3}$ isosurface).}  
    \label{fig:acoustic}
\end{figure}

Figure \ref{fig:acoustic} depicts space-time diagrams of the radial velocity along $(\theta, \phi)=(\pi/2, 0)$ (along the x-direction of Figure \ref{fig:snapshot}) for both models u8.1 and z9.6. The colors represent the magnitude of the radial velocity. The region of PNS convection, the epoch and realm of early turbulent convection, and the striations that track the emergent sound waves are all encapsulated in this plot.  The black line follows the PNS radius, defined as where the mean density equals 10$^{11}$ g cm$^{-3}$. 

However, though sound waves are unambiguously generated (Figures \ref{fig:snapshot} and \ref{fig:acoustic}), the inferred luminosity is quite small. To derive this quantity we subtract out the mean flow to obtain the pressure and speed perturbations and then use the last (sub)equation in Eq. \ref{acoustic_power} to obtain the sonic power. Figure \ref{fig:acoustic-power} demonstrates that after $\sim$0.3 seconds after bounce and just outside the PNS radius (see Figure \ref{fig:acoustic}), the emergent acoustic power is small, about five orders of magnitude below $L_{\rm PNS}$. However, at earlier times our estimate of the acoustic power emerging from deeper in gets as high as $\sim$10$^{48-49}$ ergs s$^{-1}$. The turbulence in this region at these earlier times makes it quite difficult to derive this quantity (see eq. \ref{acoustic_power} and Figure \ref{fig:acoustic}), since extracting the relevant pressure and velocity perturbations in this region at this early phase before the flow settles into a smooth, approximately spherical, wind is messy. Nevertheless, it might be that the core generated sound power reaches early on within three orders of magnitude of the peak mechanical luminosity in the PNS convection region and four orders of magnitude of $L_{\nu_e} + L_{\bar{\nu_e}}$, and that it is absorbed within a few tens of kilometers of the region of PNS convection. Wave damping would likely be due to non-linear breaking and/or neutrino cooling \citep{gossan}, and if the former could very marginally contribute to powering the supernova and subsequent wind. 

Nevertheless, our results challenge the notion that sound generated by tapping PNS convection might be a major factor in powering the supernova or in nucleosynthesis \citep{nevins2023,gossan}. However, we have yet to extract the corresponding quantities for other progenitor models. All we can say currently is that we derive quite low acoustic powers for the u8.1 and z9.6 models. It would, however, be illuminating to determine whether higher spatial resolution simulations yield a similar conclusion.

\section{Nucleosynthesis}
\label{nucleo}

The origin of the elements is the primary topic of investigation in nuclear astrophysics and has been studied in great detail for more than seventy years.
A rich literature \citep{thielemann1996,rauscher2002,woosley2002,heger2010,nomoto2013,Sukhbold2016,limongi2018,curtis2019} has been created that encompasses pre-supernova stellar evolution, explosive nucleosynthesis, $\alpha$-rich freeze-out \citep{woosley2002,nomoto2013,arcones2023,wang2023b}, post-explosion winds \citep{duncan1986,burrows1987,Burrows1995,qian1996,wanajo2018,wang2023}, and neutrino nucleosynthesis \citep{woosley1990,frohlich2006,Pruet2006} that forms the core of our general understanding of the sources of the isotopes of Nature. However, recent developments \citep{Lentz2015,wongwathanarat2015,harris2017,wanajo2018,sieverding2020,sieverding2023,wang2023,wang2023b} highlighting the differences between the traditional 1D supernova nucleosynthesis calculations with ad hoc explosions
in spherical symmetry and more modern 3D perspectives informed by self-consistent simulations of the explosion, complicated debris trajectories, and the associated nucleosynthesis are emerging to deepen, if also complicate, this narrative.    

In this context, and for this paper, we have sought to determine the differences between 1D and 3D nucleosynthetic realizations for the u8.1 and z9.6 progenitors as representatives of those (rare) models that explode in current theoretical state-of-the-art 1D simulations. We will not here be exhaustive, but will emphasize only the most salient differences we find and will defer to a later work in the context of a larger suite of progenitor models a more detailed investigation. We use the procedures outlined in \citet{wang2023b} to perform the nucleosynthetic calculations.

\begin{figure} 
    \centering
    \includegraphics[width=0.48\textwidth]{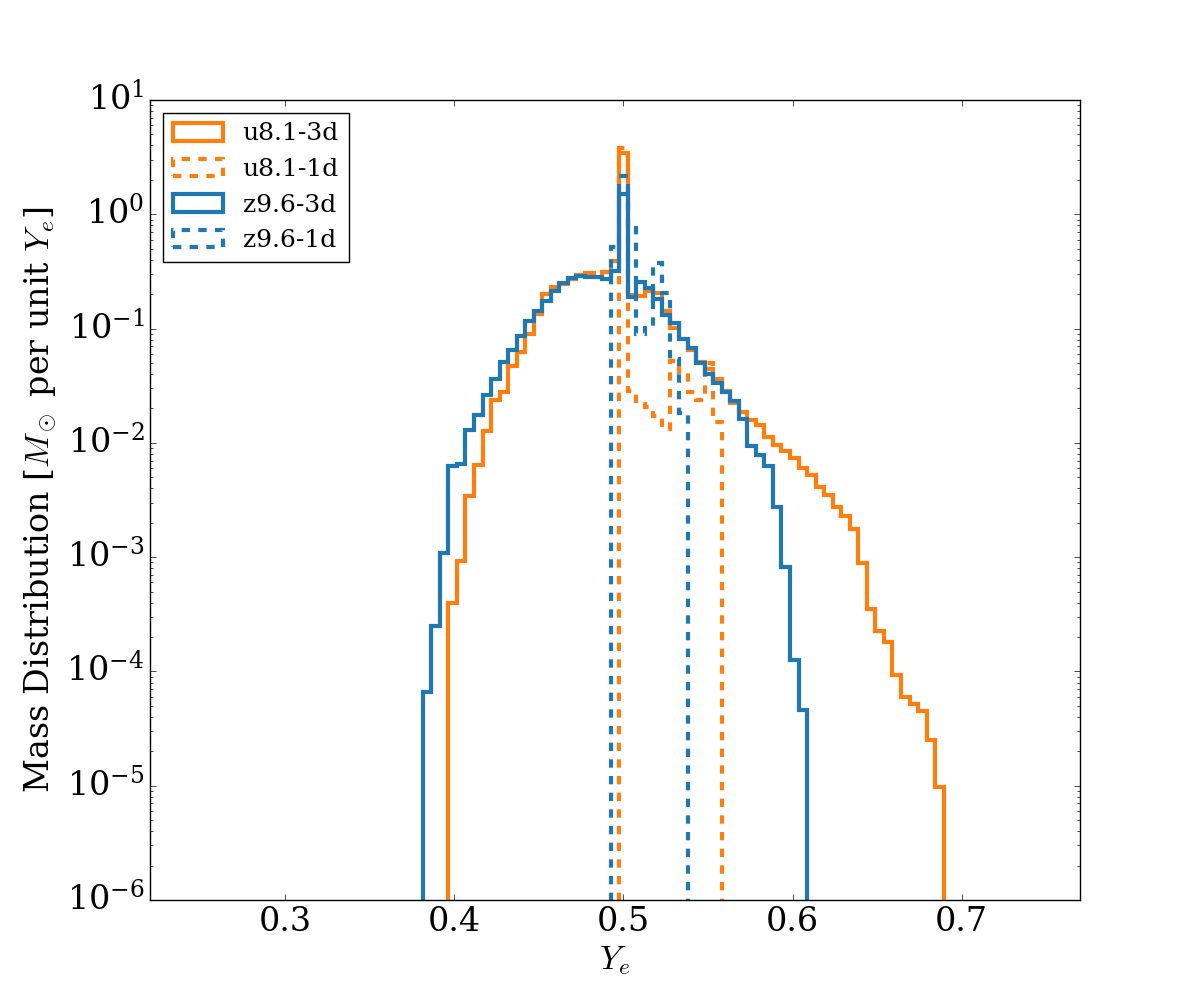}
    \includegraphics[width=0.48\textwidth]{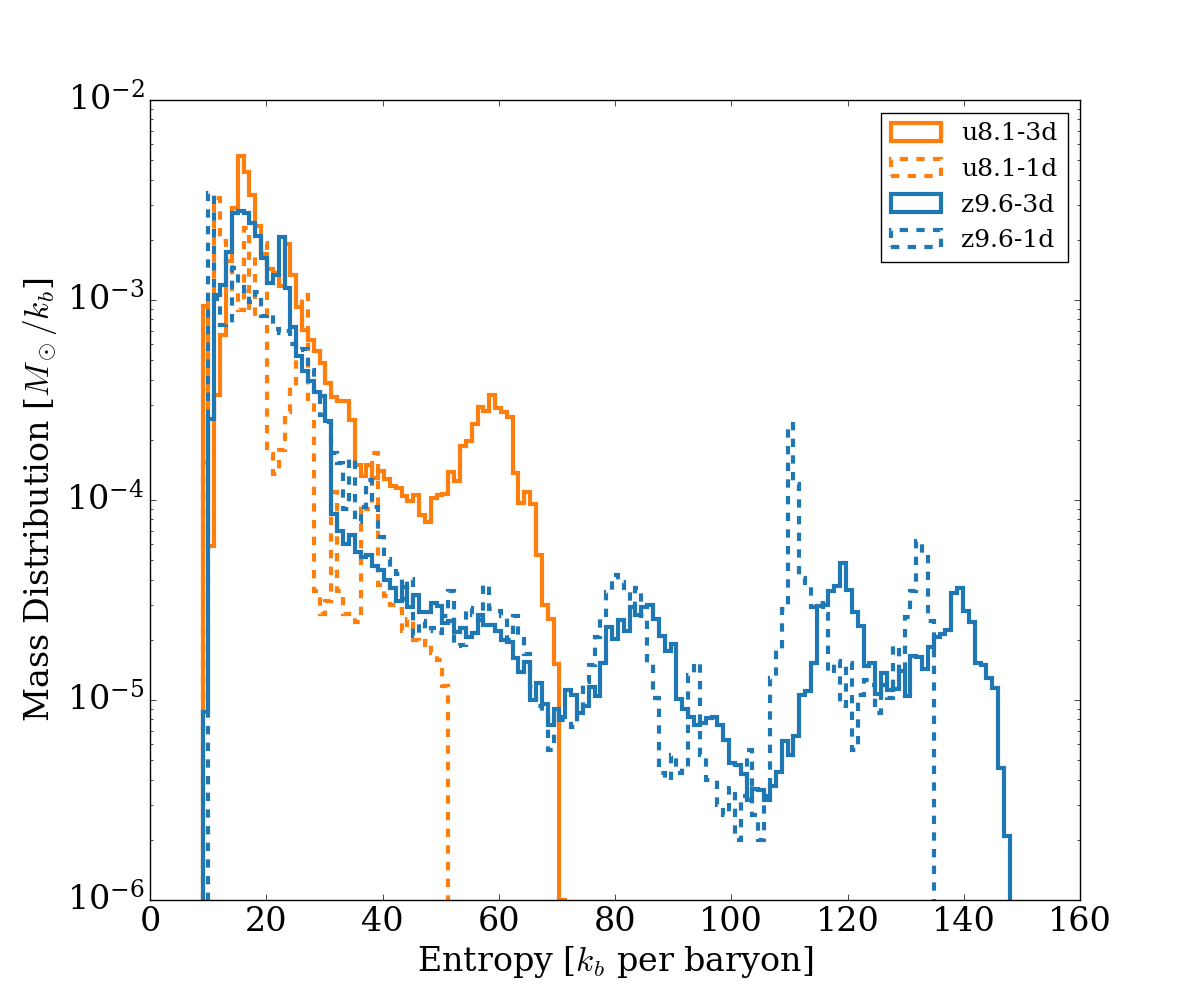}
    \caption{Electron fraction and entropy distribution of the 1D and 3D models u8.1 and z9.6. The differences in the entropy distributions of the 1D and 3D models are not large, but, due to differences in the explosion energies and mantle densities of the u8.1 and z9.6 models, differences in the associated maximum entropies emerge. Moreover, there is a tendency for the peak entropy to be slightly larger in 3D. Note that the more vigorous and earlier explosion in 3D leads to large differences in the $Y_e$ distributions with the corresponding 1D results.  In particular, the 3D ejecta contain a wider range of $Y_e$s and much more neutron-rich matter.}  
    \label{fig:tracer}
\end{figure}

The left-hand side of Figure \ref{fig:tracer} depicts the final electron fraction ($Y_e$) distributions for both the 1D (dashed) and 3D (solid) simulations of the u8.1 and z9.6 models. The differences are stark. The 1D models generally explode a bit later and less vigorously.  As a consequence, $\nu_e$ and $\bar{\nu}_e$ absorption on the ejecta has time to push the $Y_e$ into a predominantly proton-rich state, with very little neutron-rich matter ejected. In contrast, the 3D models explode earlier and more energetically, with the consequence that $\nu_e$ and $\bar{\nu}_e$ absorption does not have sufficient time to elevate initially neutron-rich matter near the PNS surface into a very proton-rich state. Moreover, the higher late-time neutrino luminosities in 3D (Figure \ref{fig:neutrino}) and the longer dwell times of a fraction of the ejecta  result in a spread in the $Y_e$ on the proton-rich side to even higher values. The net outcome is a pronounced widening of the $Y_e$ distributions of the 3D models vis \`a vis the 1D models, with significant consequences for the emergent nucleosynthesis. At the same time, the asymptotic entropy distributions (right-hand panel of Figure \ref{fig:tracer}) are not dramatically different between the 1D and 3D realizations. There are differences, however, in the entropy distributions of the u8.1 and z9.6 models themselves.

\begin{figure} 
    \centering
    \includegraphics[width=0.48\textwidth]{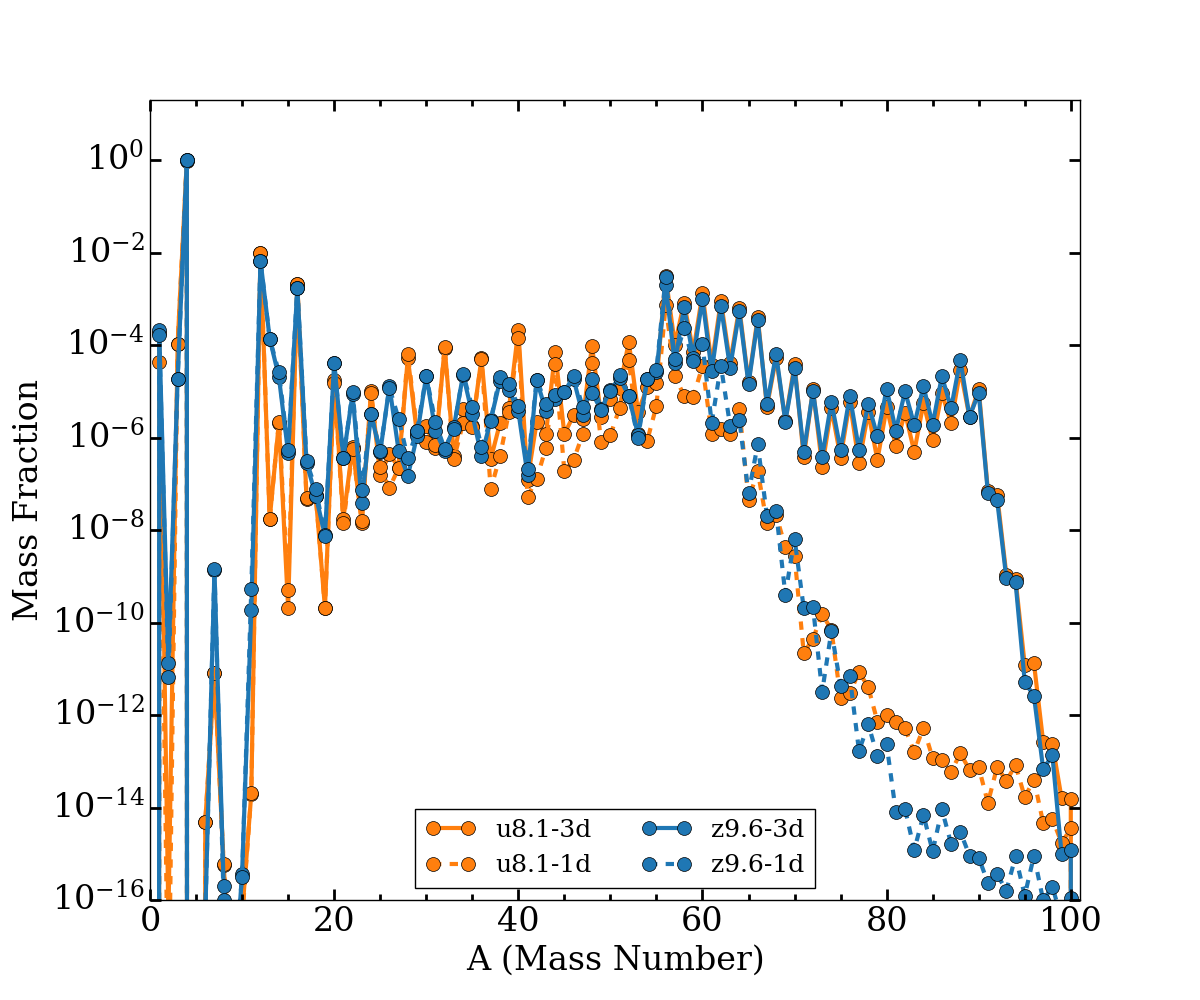}
    \includegraphics[width=0.48\textwidth]{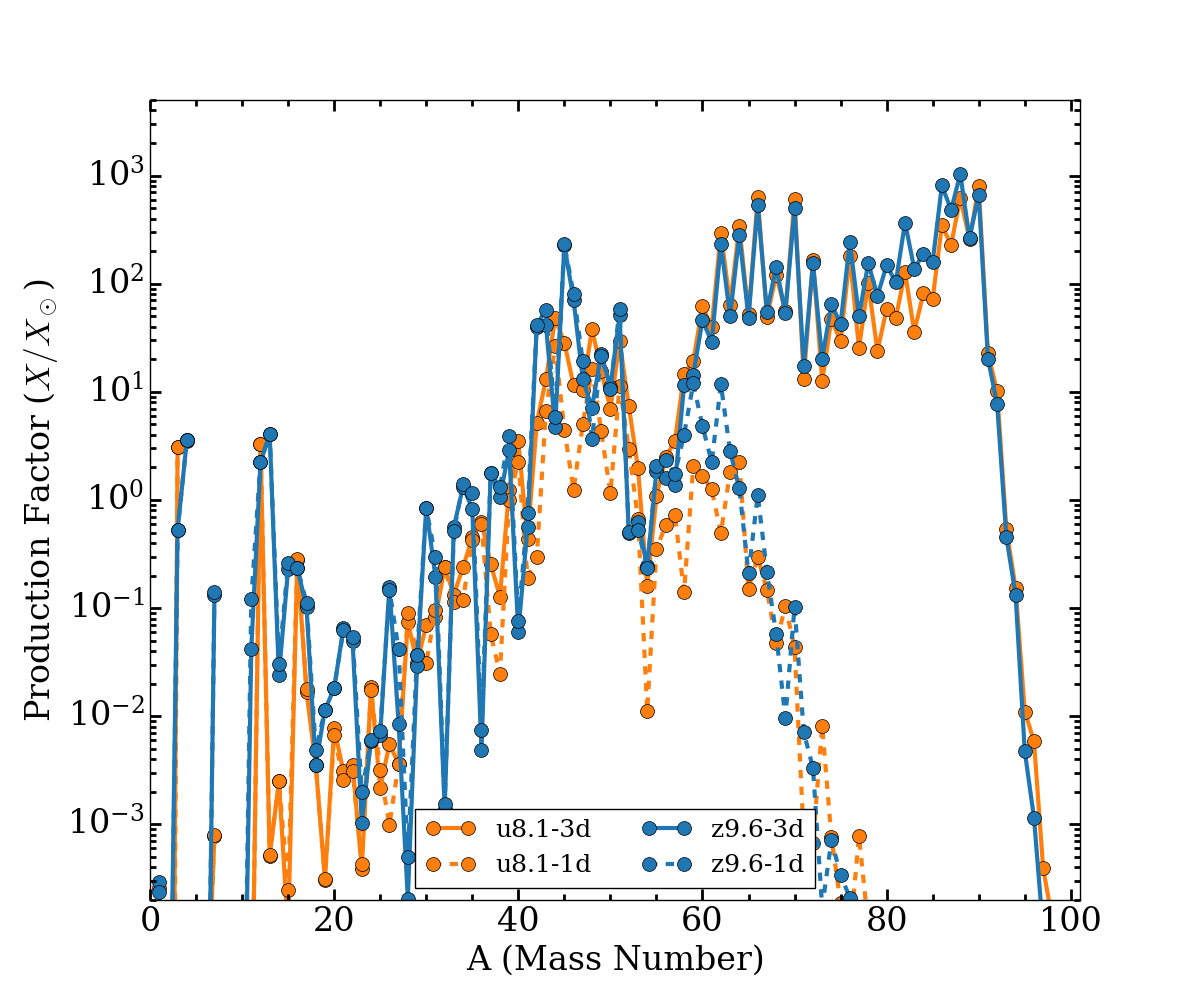}
    \caption{Ejecta abundances (left) and production factors (right) for the 1D (dashed) and 3D (solid) u8.1 (orange) and z9.6 (blue) progenitors. Note that only the 3D models manifest significant ``weak" r-process production near $A=90$.}  
    \label{fig:abundance}
\end{figure}

Figure \ref{fig:abundance} shows one of the major consequences.  On the left hand panel is the mass fraction distribution of the ejecta versus atomic mass number ($A$). On its right-hand panel is the corresponding production factor (see \citet{wang2023b}). For the lower $A$s, most of the nucleosynthesis is accomplished prior to the supernova during stellar evolution, and the models are quite similar. However, for the higher $A$s on these plots there is a pronounced difference in the yields. The 1D models do not generate much near the first peak of the r-process (a ``weak" r-process), while the 3D models clearly do, and this is a direct result of the presence of significant neutron-rich ejecta in them.  This conclusion echoes a similar conclusion in \citet{wang2023} concerning the nucleosynthetic differences between the s9.0a and s9.0b models. The former (with imposed initial perturbations that kick start turbulence) experienced a more rapid explosion that left enough neutron-rich ejecta to result in significant weak r-process production\footnote{We note that \citet{wang2023} did not find that imposed perturbations made much of a nucleosynthetic difference for those models with a greater post-bounce delay before explosion.}. These results (re)emphasize the potential importance of the lower mass CCSN progenitors studied in \citet{wang2023} and \citet{wang2023b} as contributing sources of the first peak of the r-process. We provide the corresponding $^{56}$Ni yields in Table 1. 

Curiously, the $^{48}$Ca yields for the 3D u8.1 and z9.6 models are $\sim$1.58$\times$10$^{-5}$ $M_{\odot}$ and $\sim$3.31$\times$10$^{-5}$ $M_{\odot}$, respectively. This reflects the neutron-richness (Figure \ref{fig:tracer}) of a large fraction of their ejecta and emphasizes, as do the results for the s9.0a model, the possibility that this most neutron-rich of stable nuclei might originate in the explosions of the lowest mass massive stars. The corresponding $^{44}$Ti yields are $\sim$3.5$\times$10$^{-6}$ $M_{\odot}$ and $\sim$2.9$\times$10$^{-6}$ $M_{\odot}$, respectively, with the values still growing at the end of the simulation (Wang and Burrows, in preparation). This emphasizes again the important role played by the neutrino-driven winds in $^{44}$Ti production, as discussed in \citet{sieverding2023,wang2023b}.

For $^{57}$Ni, the yields are $\sim$1.5$\times$10$^{-4}$ $M_{\odot}$ and $\sim$1.4$\times$10$^{-4}$ $M_{\odot}$ for \tianshu{3D} models u8.1 and z9.6, respectively, resulting in $^{56}$Ni to $^{57}$Ni ratios of $\sim$40.  This is near the solar value of $\sim$43. Moreover, the stable Ni ($^{58}$Ni, $^{60}$Ni, and $^{62}$Ni) to $^{56}$Ni ratio of the ejecta is around $\sim$1.0$-$1.5 for both 3D models, or 15$-$30 relative to the solar ratio of $\sim$0.057 \citep{karrer2007}. If the u8.1 and z9.6 models had had solar metallicity envelopes, the ratio of the aggregate nickel to iron ejecta (after $^{56}$Ni decay) would have been $\sim$5$-$6, and this may be what we might expect for
a solar-metallicity model that theoretically would have exploded in 1D.  However, such a conclusion must await a proper simulation of the relevant progenitor.


\section{Conclusions}   
\label{conclusion}



In this paper we have focused on the behavior of the cores of representative massive stars that technically explode in spherical symmetry. According to current theory, this is a small subclass of supernova progenitors with very tenuous mantles that at solar-metallicity might have ONe cores at the onset of collapse and inhabit the $\sim$8$-$9 $M_{\odot}$ mass range \citep{Nomoto1984,limongi2024}. These would be so-called ``electron-capture" supernovae. The classic such model is the solar-metallicity 8.8 $M_{\odot}$ model of \citet{Nomoto1984}, which was shown originally to explode in 1D by \citet{Kitaura2006} and in 2D by \citet{burrows2007_after}. However, as the behavior of models u8.1 and z9.6 demonstrates, the mass range at extremely low metallicity for which models theoretically explode in 1D may be broader and such progenitors may in fact have iron cores at the onset of collapse. 

When simulated in 3D with a modern supernova simulation code, we find that the observable outcomes of representative progenitors of this class (for this paper chosen to be the u8.1 and z9.6 models of A. Heger) are quite different than found in 1D in almost all relevant quantities.  In particular, we find for the u8.1 and z9.6 progenitors that in 3D the explosion energies are 0.171 B and 0.207 B and the $^{56}$Ni yields are 0.572$\times$10$^{-2}$ $M_{\odot}$ and 0.566$\times$10$^{-2}$ $M_{\odot}$, while in 1D the corresponding quantities are 0.017 B and 0.129 B and 0.164$\times$10$^{-2}$ $M_{\odot}$ and 0.922$\times$10$^{-2}$ $M_{\odot}$ (see Table 1). The large differences in the explosion energy and $^{56}$Ni yield between the 3D and 1D simulations of the u8.1 progenitor reflect in part the significant delay to explosion of the latter, associated with its slightly denser mantle (Figure \ref{fig:M-rho}). However, the explosion energies are significantly larger in 3D in both progenitors, a consequence not only of the shorter delay to explosion (most prominant for the u8.1 model), but also of the presence in the core of PNS convection (suppressed in 1D). Such lepton-driven convection \citep{Dessart2006,Nagakura2020} boosts the driving neutrino luminosities by as much as $\sim$50\% at later times (Figure \ref{fig:neutrino})
\footnote{We note that \citet{melson2015b} performed a 3D simulation of the same z9.6 model and obtained an explosion energy of 0.08$-$0.1 B by the end of their simulation (taken to 0.4 seconds after bounce). \citet{Stockinger2020} also simulated the z9.6 model in 3D and obtained an explosion energy of $\sim$0.086 B. These explosion energies are a factor of $\sim$two lower than what we obtain and the detailed origin of this difference is unknown. Both \citet{melson2015b} and \citet{Stockinger2020} used the LS220 nuclear EOS (not the SFHo EOS), but we think it unlikely that this alone could explain the difference. Slight differences in the neutrino-matter coupling rates 
might translate non-linearly into a quantitative difference in the asymptotic energy. Whatever the reason, it would be useful to determine the origin of this quantitative difference.}. The range of final gravitational masses is narrow, and for the 3D models is $\sim$1.25 $M_{\odot}$.  This number reflects not only the initial stellar model core structure, but the nuclear equation of state (SFHo) employed in our simulations.

We find that the distribution of the electron fraction ($Y_e$) of the debris is significantly broadened for the 3D models with respect to their 1D counterparts. The more neutron-rich 3D ejecta contain significant weak r-process and $^{48}$Ca yields that may be important features of this lower end of the CCSN progenitor mass range. This has also been seen in the solar metallicity s9.0a model \citep{wang2023b,wang2023}. In addition, we find that the stable nickel to iron ratio of these models is $\sim$20, and is much higher than that measured for the Crab Nebula (T. Temim, private communication).  However, the corresponding ratio for our lower-mass/solar-metallicity/iron-core progenitor models s9.0a and s11 do comport with the Crab measurements, though the ratios for most of the rest of our recent 3D simulation suite \citep{burrows2024} are roughly solar. What this may mean will be discussed in Wang \& Burrows (in preparation). 

With such tenuous mantles, neutrino-driven mechanical winds emerge quickly and are a major factor in the overall supernova energy. Importantly, after a few tenths of a second after bounce, \tianshu{$L_{\rm mech}$} is $\sim$0.25 and $\sim$0.15 times smaller in 1D than 3D for the u8.1 and z9.6 progenitors, respectively. 
We also find that $L_{mech} \propto \dot{Q}^{1.8-1.6}$, where $\dot{Q}$ is the total neutrino energy deposition rate at the base of the wind. 

Coupled with the results for a much broader spectrum of progenitors \citep{burrows2024}, the results we witness in this paper support the idea that there is a continuity in dynamical behavior and outcomes across the entire progenitor continuum and that the lower mass realm is not qualitatively distinct in terms of the explosion mechanism or dynamics.  However, there may be an interesting stellar metallicity dependence of the explosion energy at the low mass end, with lower metallicity explosions for these progenitors having $\sim$twice the explosion energy at the same low ZAMS mass. This conclusion, however, has yet to be made robust and will require further study. 

Interestingly, we find that the cores of the 3D models (as opposed to 1D models) radiate acoustic power that may play a modest, though quite sub-dominant, role in supernova and wind dynamics \citep{gossan,nevins2023}. This component seems sourced in the fluctuations associated with inner PNS convection and has a frequency range between $\sim$100 and $\sim$250 Hz. Whether acoustic power plays an important role anywhere across the supernova progenitor continuum is unlikely, but this has yet to be investigated in adequate detail. 

In conclusion, we have found that though a model might be found theoretically to explode in 1D (spherical symmetry), one must perform modern supernova simulations in 3D to capture most of the associated observables. The differences between 1D and 3D models are just too large to ignore.

\section*{Acknowledgments}

We thank Tea Temim, David Vartanyan, Chris White, and Matt Coleman for previous conversations and/or ongoing collaborations and Alex Heger for sharing his u8.1 and z9.6 progenitor models.  We acknowledge support from the U.~S.\ Department of Energy Office of Science and the Office of Advanced Scientific Computing Research via the Scientific Discovery through Advanced Computing (SciDAC4) program and Grant DE-SC0018297 (subaward 00009650) and support from the U.~S.\ National Science Foundation (NSF) under Grants AST-1714267 and PHY-1804048 (the latter via the Max-Planck/Princeton Center (MPPC) for Plasma Physics). Some of the models were simulated on the Frontera cluster (under awards AST20020 and AST21003), and this research is part of the Frontera computing project at the Texas Advanced Computing Center \citep{Stanzione2020}. Frontera is made possible by NSF award OAC-1818253. Additionally, a generous award of computer time was provided by the INCITE program, enabling this research to use resources of the Argonne Leadership Computing Facility, a DOE Office of Science User Facility supported under Contract DE-AC02-06CH11357. Finally, the authors acknowledge computational resources provided by the high-performance computer center at Princeton University, which is jointly supported by the Princeton Institute for Computational Science and Engineering (PICSciE) and the Princeton University Office of Information Technology, and our continuing allocation at the National Energy Research Scientific Computing Center (NERSC), which is supported by the Office of Science of the U.~S.\ Department of Energy under contract DE-AC03-76SF00098.

\section*{Data Availability}

The summary numerical data underlying this article will be shared upon reasonable request to either author.



\bibliography{citations}{}
\bibliographystyle{aasjournal}

\end{document}